\documentclass[%
 reprint,
 amsmath,amssymb,
 aps,
]{revtex4-2}

\usepackage{graphicx, subfigure, float}
\usepackage{dcolumn}
\usepackage{bm}

\begin{document}

\preprint{APS/123-QED}

\title{Particle production during Inflation with a non-minimally coupled spectator scalar field}

\author{Zhe Yu}
\email{yuzhe@ihep.ac.cn}
\affiliation{Key Laboratory for Particle Astrophysics, Institute of High Energy Physics, Chinese Academy of Sciences, 19B Yuquan Road, Beĳing 100049, China}
\affiliation{School of Physical Sciences, University of Chinese Academy of Sciences, No.19A Yuquan Road, Beijing 100049, China
}

\author{Chengjie Fu}
\email{fucj@ahnu.edu.cn}
\affiliation{Department of Physics, Anhui Normal University, Wuhu, Anhui 241000, China}
\author{Zong-Kuan Guo}
\email{guozk@itp.ac.cn}
\affiliation{CAS Key Laboratory of Theoretical Physics, Institute of Theoretical Physics,
Chinese Academy of Sciences, P.O. Box 2735, Beijing 100190, China
}
\affiliation{School of Physical Sciences, University of Chinese Academy of Sciences, No.19A Yuquan Road, Beijing 100049, China
}
\affiliation{School of Fundamental Physics and Mathematical Sciences, Hangzhou Institute for Advanced Study, University of Chinese Academy of Sciences, Hangzhou 310024, China}

\date{\today}

\begin{abstract}
We study the inflationary model with a spectator scalar field $\chi$ coupled to both the inflaton and Ricci scalar. The interaction between the $\chi$ field and the gravity, denoted by $\xi R\chi^2$, can trigger the tachyonic instability of certain modes of the $\chi$ field. As a result, the $\chi$ field perturbations are amplified and serve as a gravitational wave (GW) source. When considering the backreaction of the $\chi$ field, an upper bound on the coupling parameter $\xi$ must be imposed to ensure that inflation does not end prematurely. In this case, we find that the inflaton's evolution experiences a sudden slowdown due to the production of $\chi$ particles, resulting in a unique oscillating structure in the power spectrum of curvature perturbations at specific scales. Moreover, the GW signal induced by the $\chi$ field is more significant than primordial GWs at around its peak scale, leading to a noticeable bump in the overall energy spectrum of GWs. It's worth noting that this bump predicted in the slow-roll inflationary scenario is unlikely to be detected by LISA and Taiji, but there is a slim chance it might approach the detection limits of GW experiments like BBO and SKA if we devise distinctive inflatonary potentials.
\end{abstract}

\maketitle


\section{\label{sec:intro}Introduction}

The inflationary scenario \cite{Guth:1980zm,Starobinsky:1980te,Sato:1980yn,Linde:1981mu} has become the dominant paradigm of the early Universe to address the horizon and flatness problems in the standard cosmology. During inflation, quantum vacuum fluctuations are stretched out to the super-horizon scales and become primordial perturbations \cite{Mukhanov:1981xt,Sasaki:1983kd,Kodama:1984ziu}, where the scalar modes (i.e., primordial curvature perturbations) result in the observed cosmic microwave background (CMB) anisotropies and the large-scale structures (LSS). Thanks to the accurate CMB and LSS measurements, the amplitude of the power spectrum for curvature perturbations has been precisely constrained as $2.1\times 10^{-9}$ at $k=0.05\rm{Mpc}^{-1}$ with a slight scale dependence \cite{Planck:2018jri}, which is consistent with the prediction of the general single-field slow-roll inflation. Moreover, an extremely important prediction of inflation is the generation of a stochastic background of primordial gravitational waves (GWs), characterized by a nearly scale-invariant power spectrum. By the observations of CMB B-mode polarization, the current bound on the tensor-to-scalar ratio $r$, describing the amplitude of primordial GWs, has been found to be $r<0.036$ at $95\%$ confidence level for $k=0.05\rm{Mpc}^{-1}$ \cite{BICEP:2021xfz}.

Although one can obtain information about the inflationary physics by the observations of primordial perturbations, the CMB only probe a small fraction of inflation associated with the large scales ($k\lesssim 1 \rm{Mpc}^{-1}$). However, the GW detection open a new window to observe primordial perturbations at smaller scales to shed light on the picture of the last stages of inflation. The ongoing and planned GW experiments such as pulsar timing array (EPTA\cite{EPTA:2023fyk,EPTA:2023gyr,EPTA:2023xxk},NANOGrav \cite{NANOGrav:2015aud,NANOGRAV:2018hou}, SKA \cite{Carilli:2004nx,Janssen:2014dka}),  ground-based interferometers (LIGO~\cite{Harry:2010zz}, Virgo~\cite{VIRGO:2014yos}), and space-based interferometers ( LISA~\cite{LISA:2017pwj}, Taiji~\cite{Ruan:2018tsw}) have the potential to detect the stochastic GW background in the range of frequencies between the nHz and kHz range, covering scales around $10^{6}$-$10^{18}\mathrm{Mpc}^{-1}$. However, current bounds from CMB observations predict primordial GWs, originating from quantum vacuum fluctuations within the general single-field slow-roll framework, to be out of reach for these experiments due to the nearly scale-invariance of the GW spectrum, whose amplitude is suppressed at the small scales. Nevertheless, the possibility of detecting the GW background from inflation through these experiments cannot be dismissed, especially if some specific inflationary models produced a GW signal with a large amplitude and a significant deviation from scale-invariance \cite{Guzzetti:2016mkm,Bartolo:2016ami}. During inflation, GWs can be generated through a classical mechanism in which the equation of motion for GWs incorporates a source term. Such a term emerges if additional fields, present during inflation, have interactions with the inflaton resulting in strong particle production, such as the gauge particle production through the coupling of the pseudo-scalar inflaton to gauge fields \cite{Barnaby:2010vf,Sorbo:2011rz,Barnaby:2011vw,Barnaby:2011qe,Anber:2012du,Ferreira:2014zia,Domcke:2016bkh,Peloso:2016gqs,Cheng:2018yyr,Ozsoy:2020kat}.

In this paper, we focus on the situation of the scalar particle production during inflation, which has been widely studied in the literatures~\cite{Romano:2008rr,Barnaby:2009mc,Green:2009ds,Cook:2011hg,Carney:2012pk,Fedderke:2014ura,Ozsoy:2014sba,Goolsby-Cole:2017hod}. It is a simple way to achieve such a situation by introducing an extra scalar field $\chi$ that interacts with the inflaton $\phi$ via the coupling \cite{Barnaby:2009mc,Cook:2011hg}
\begin{align}
\label{interaction}
   \frac{g^2}{2}\left(\phi-\phi_0\right)^2 \chi^2,
\end{align}
where $\phi_0$ is a constant having the dimension of mass, and $g$ denotes a dimensionless coupling constant. The effective mass of the $\chi$ field, $m_\chi=g\left|\phi-\phi_0\right|$, is related to the value of the inflaton $\phi$, and vanishes exactly when $\phi=\phi_0$. For a short period when the inflaton crosses around $\phi_0$, the mass $m_\chi$ changes non-adiabatically such that the specific momentum modes of the $\chi$ field are excited and act as a classical source of GWs. However, it has been pointed out that the production of quanta of an extra scalar field interacting with the inflaton as described by Eq.\eqref{interaction} induces an insignificant GW signal compared with primordial GWs \cite{Cook:2011hg,Guzzetti:2016mkm}. When taking into account that the $\chi$ field becomes massless during inflation due to its coupling with another scalar field (other than the inflaton), the resulting spectrum for induced GWs does increase, however it remains significantly smaller compared to the spectrum of primordial GWs and does not dominate in terms of overall contribution \cite{Goolsby-Cole:2017hod}. In this paper, we explore a scenario where the $\chi$ field is coupled to the Ricci scalar $R$ through the $\xi R \chi^2$ term (with $\xi$ being dimensionless coupling parameter) in addition to the interaction term \eqref{interaction}. In this scenario, during a brief period when the inflaton traverses  through $\phi_0$, the effective mass square of the $\chi$ field become negative as a result of the non-minimal coupling of the $\chi$ field to gravity. Consequently, the $\chi$ field undergoes a tachyonic instability leading to an irruptive production of $\chi$ particles. This scenario proves to be more efficient in generating GWs compared to the case with minimal coupling. 

Our paper is organized as follows. In section \ref{sec2}, we start by introducing the inflationary model, where a spectator scalar field $\chi$ is coupled to both the inflaton and the Ricci scalar. We then investigate the amplification of the $\chi$ field due to the tachyonic instability, and the production of GWs induced by the $\chi$ field. In section \ref{sec3}, we turn our attention to the phenomenology of this inflationary scenario in light of the backreaction of the amplified $\chi$ field on the background and perturbation evolution. In section \ref{sec4}, we carefully study the testability of these GWs by analytical method. Finally, the conclusion and discussion are given in section \ref{sec5}. Throughout the paper, we adopt $c=\hbar=1$ and the reduced Planck mass defined as $M_{\mathrm{p}}=1/\sqrt{8\pi G}$. 

\section{\label{sec2}Model}
Our model incorporates an extra scalar field that is coupled to both the inflaton and Ricci scalar, within the framework of general single-field slow-roll inflation. This is specified by the following action,
\begin{widetext}
\begin{align}
\label{action}
S=\int d^4 x \sqrt{-g}\left[\frac{M_{\mathrm{p}}^2}{2}R-\frac{1}{2}\nabla^\mu \phi  \nabla_\mu \phi-V(\phi)-\frac{1}{2} \nabla^\mu\chi\nabla_\mu\chi-\frac{g^2}{2}\left(\phi-\phi_0\right)^2 \chi^2+\frac{1}{2}\xi R \chi^2\right],
\end{align}
\end{widetext}
where the $\phi$ field serves as the inflaton, and the $\chi$ field is a spectator scalar field. Note that in this study, we do not consider an initial homogeneous background for $\chi$ field. Therefore, we treat the $\chi$ field as a quantum field. In this section, our primary focus lies on investigating the efficiency of the GW generation resulting from $\chi$-particle production. Hence, we disregard the inflaton perturbations, the metric perturbations, and the backreaction caused by $\chi$ particles.

The equation of motion for the $\chi$ field is given by,
\begin{align}
\label{EoM_chi}
\ddot{\chi}+3H\dot{\chi}-\frac{1}{a^2}\nabla^2\chi+\left[g^2\left(\phi-\phi_0\right)^2-\xi R\right]\chi=0,
\end{align}
where $R=6(\dot H + 2H^2)$. Then, the quantum field $\chi$ can be decomposed as
\begin{align}
\label{decomposion_chi}
\chi=\frac{1}{(2\pi)^{3/2}} \int d^3k\left[\chi_k\hat{a}_{\boldsymbol{k}} + \chi^\ast_{-k} \hat{a} ^\dagger_{-\boldsymbol{k}}  \right]e^{i \boldsymbol{k} \cdot \boldsymbol{x}},
\end{align}
where the creation and annihilation operators $\hat{a}^\dagger_{\boldsymbol{k}}$ and $\hat{a}_{\boldsymbol{k}}$ satisfy the canonical commutation relation $[\hat{a}_{\boldsymbol{k}},\hat{a}^\dagger_{\boldsymbol{k}^\prime}]=\delta(\boldsymbol{k}-\boldsymbol{k}^\prime)$. The mode functions $\chi_k$ obey the following equation of motion,
\begin{align}
\label{EoM_chi_k}
\ddot{\chi}_k+3H \dot{\chi}_k+\omega_k^2 \chi_k =0,
\end{align}
with
\begin{align}
\label{omega_1}
\omega_k^2=\frac{k^2}{a^2}+g^2\left(\phi-\phi_0\right)^2-6\xi(\dot H + 2H^2),
\end{align}
which reduces to the following form,
\begin{align}
\label{omega_2}
\omega_k^2 \simeq \frac{k^2}{a^2}+g^2\left(\phi-\phi_0\right)^2-12\xi H^2,
\end{align}
under the slow-roll approximation during inflation. In a specific range of parameter values with $\xi>0$, we can observe the following intriguing phenomena: as the inflaton rolls down the potential up to around value of $\phi_0$, the contribution of the $g^2$-term in Eq. \eqref{omega_2} becomes negligible, which gives rise to a scenario where $\omega_k^2<0$ for certain cases. Consequently, the modes with $k/(aH)<\sqrt{12\xi}$ will experience a tachyonic instability until the inflaton moves far away from $\phi_0$, resulting in an amplification of the corresponding mode functions. Meanwhile, these modes serve as a source term in the GW equation of motion. Next, we will numerically investigate the amplification of the vacuum fluctuations of $\chi$ and the generation of induced GWs.

\begin{figure}
\centering
\includegraphics[width=0.38\textheight]{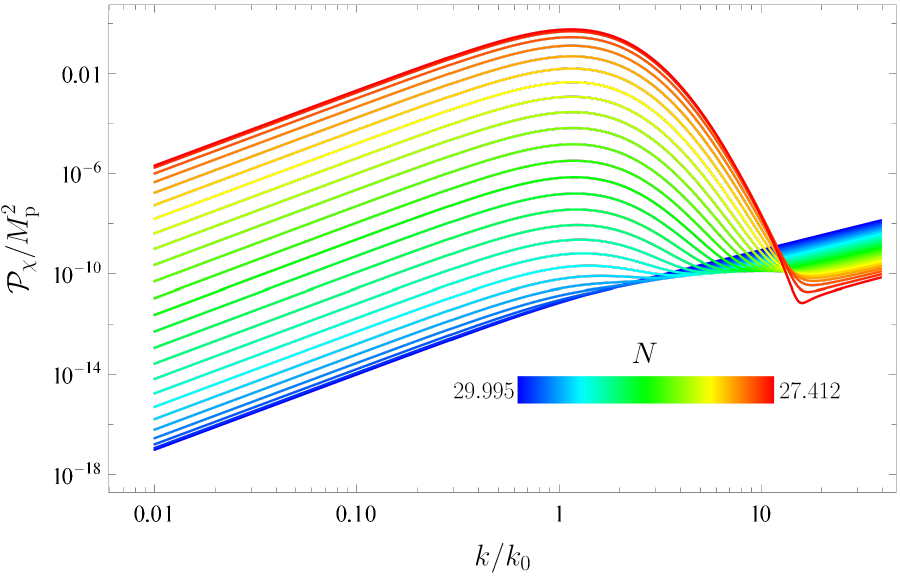}
\includegraphics[width=0.38\textheight]{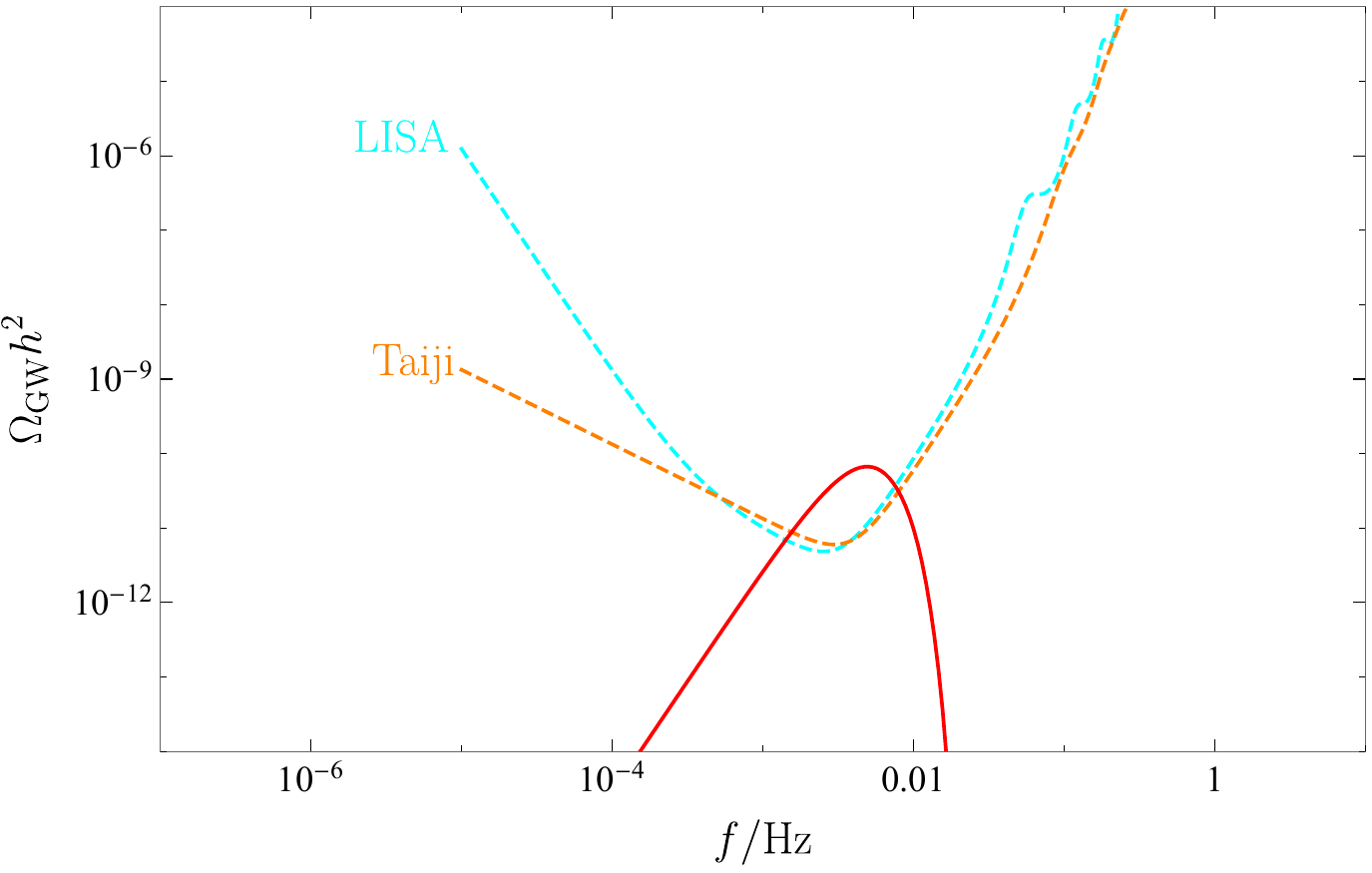}
\caption{{\it Top panel}: The evolution of the power spectrum $\mathcal{P}_\chi=k^3/(2\pi^2)|\chi_k|^2$ as a fucntion of $k/k_0$ with the {\it e}-folding number $N$. Here, $k_0$ ($=a_0H_0$) denotes the scale exiting the horizon at the time when $\phi=\phi_0$, and $N$ is defined as $N=\ln(a_e/a)$ with $a_e$ being the scale factor at the end of inflation. {\it Bottom panel}: The present energy spectrum of induced GWs. This prediction is obtained by setting {\it e}-folds $N$ at the time when the CMB scale $k=0.05\rm{Mpc}^{-1}$ exits the horizon as $60$, which is adopted throughout the paper. The cyan and orange dashed lines represent the sensitivity curves of LISA and Taiji, respectively.}
\label{fig1}
\end{figure}

To illustrate the results for this model, we adopt the Starobinsky potential, i.e., $V(\phi)=M^2M_{\mathrm{p}}^2 \left[1-\exp\left(-\sqrt{2/3}~ \phi/{M_{\mathrm{p}}}\right)\right]^2$ with $M=9.53 \times 10^{-6}M_{\mathrm{p}}$, as a typical representative. Within the spatially flat FRW metric, the inflationary dynamics is determined by the following field equations, 
\begin{align}\label{Ein_Eq_wo_Br}
-2\dot H - 3H^2= M_{\mathrm{p}}^2 \left( \frac{1}{2}\dot\phi^2 - V \right),
\end{align}
\begin{align}\label{KG_Eq_wo_Br}
\ddot{\phi}+3 H \dot{\phi}+ \frac{dV}{d\phi}=0,
\end{align}
Then, we numerically solve the coupled set of background equations \eqref{Ein_Eq_wo_Br} and \eqref{KG_Eq_wo_Br} and perturbation equation \eqref{EoM_chi_k} using the initial Bunch-Davies vacuum state described by
\begin{align}
\chi_k = \frac{1}{a\sqrt{2k}}, \quad \dot{\chi}_k=-i\frac{k}{a}\chi_k-H\chi_k.
\end{align}
In the top panel of Fig. \ref{fig1}, we plot the evolution of the power spectrum of the $\chi$ field from $N_1=29.995$ to $N_2=27.412$ for the parameter set of $\phi_0=4.57M_{\mathrm{p}}$, $g = 100 M/M_{\mathrm{p}}$ and $\xi=6$. Here, $N_1$ represents the {\it e}-folding number at the onset of the tachyonic instability, while $N_2$ corresponds to the {\it e}-folding number at which the tachyonic instability ends basically and the power spectrum reaches its maximum. It is easy to observe that during this phase, the modes with $k\lesssim 10k_0$, which is well agreement with our foregoing estimation $k/(a_0H_0)<\sqrt{12\xi}$, are significantly amplified due to the tachyonic instability. The intensity of the instability increases as $k$ decreases. However, it is noteworthy that each mode shares same intensity of instability when $k<a_0H_0 \ll \sqrt{12\xi}a_0H_0$, where the $\xi$-term in Eq. \eqref{omega_2} dominates over other terms. As a result, the power spectrum $\mathcal{P}_\chi$ features a peak at around $k_0$ and has a $k^3$ slope in the infrared region. For modes with $k > 10 k_0$, since $\omega_k^2>0$ still holds true when the inflaton reaches around $\phi_0$, the amplitude of the corresponding power spectrum decays with the expansion of the Universe.

By numerically computing the integral given in Eq. \eqref{PS_GW} and utilizing the relation \eqref{ES_GW}, we can determine the present energy spectrum of GWs induced by the amplified modes of the $\chi$ field, which is displayed in the bottom panel of Fig. \ref{fig1}. It is evident that the GW energy spectrum exhibits a peak at frequencies within the range detectable by LISA and Taiji, surpassing their sensitivity curves. This indicates that the $\chi$ field serves as an extremely efficient source of GWs in our model. However, it is important to note that these results are obtained under the assumption of neglecting the backreaction of the $\chi$ field. 

\begin{figure}
\centering
\includegraphics[width=0.38\textheight]{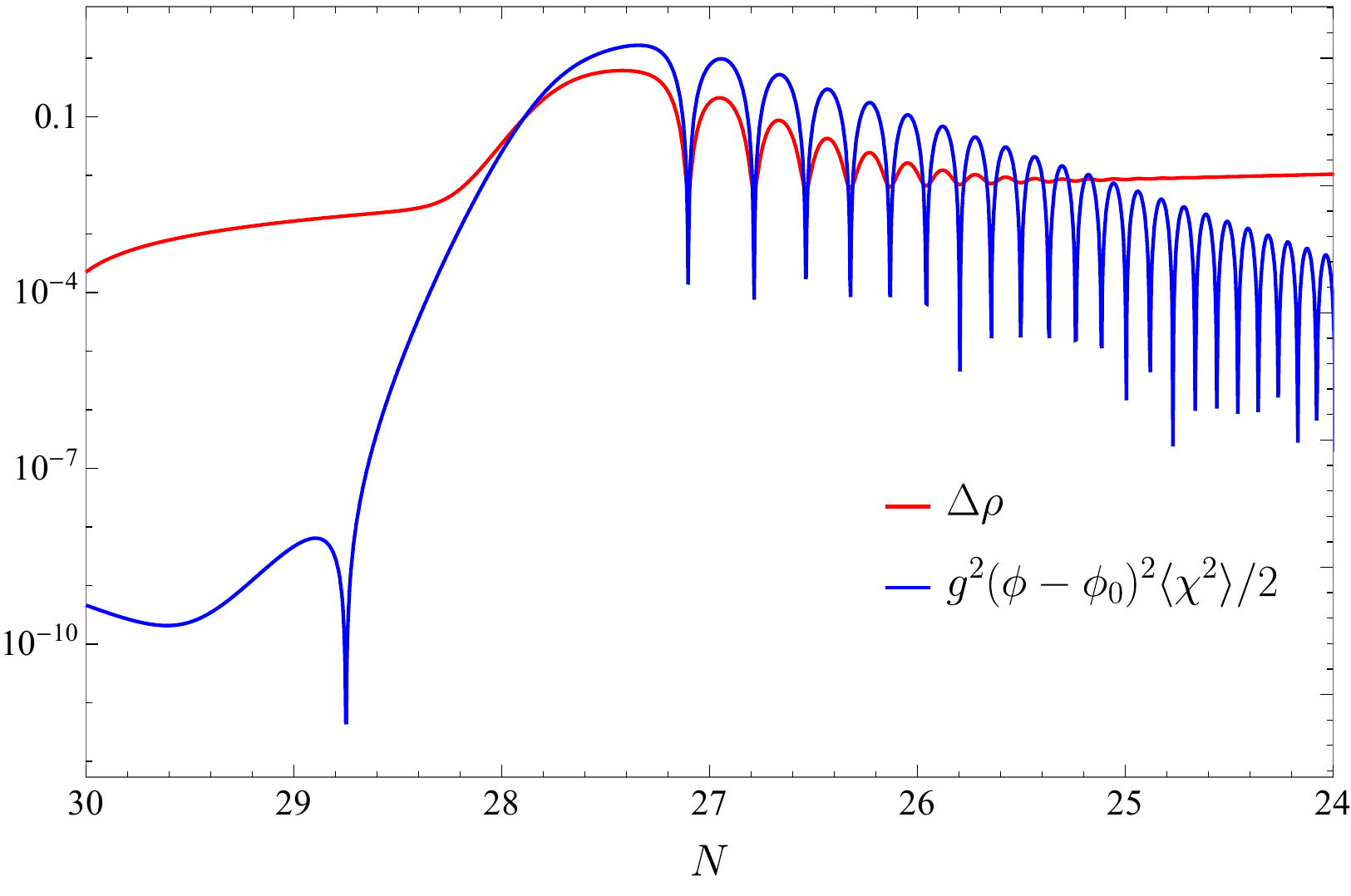}
\caption{The evolution of $\Delta \rho$ and $g^2\left(\phi-\phi_0\right)^2\left\langle\chi^2\right\rangle/2$ as a function of the {\it e}-folding number $N$.}
\label{fig2}
\end{figure}

It has been pointed out in Ref. \cite{Inomata:2021zel} that the energy conservation law imposes an upper bound on the energy density of the amplified field fluctuations. Specifically, it can be expressed as follows:
\begin{align}\label{ECL}
\rho_f(N) < \Delta\rho(N) = \frac{1}{2}\dot\phi(N_1)^2+  V(\phi(N_1)) - \rho_{\rm pot}(N)
\end{align}
where the energy density of the $\chi$ field fluctuations $\rho_f$ and the potential energy density of the inflaton $\rho_{\rm pot}$ can be expressed, based on Eq. \eqref{Ein_Eq1_w_Br}, as follows
\begin{widetext}
\begin{align}\label{rho_f}
\rho_f(N<N_1) = \frac{1}{1+\xi M_{\rm{p}}^{-2}\left\langle\chi^2\right\rangle}&\left[\frac{1}{2}\left\langle\dot{\chi}^2\right\rangle+\frac{1}{2 a^2}\left\langle(\nabla\chi)^2\right\rangle+\frac{1}{2}g^2\left(\phi-\phi_0\right)^2\left\langle\chi^2\right\rangle -6\xi H\left\langle\chi\dot\chi\right\rangle+\frac{\xi}{a^2}\left\langle\nabla^2(\chi^2) \right\rangle\right],
\end{align}
\end{widetext}
\begin{align}
\rho_{\rm pot}(N<N_1) = \frac{V(\phi)}{1+\xi M_{\rm{p}}^{-2}\left\langle\chi^2\right\rangle}.
\end{align}
Since the $g^2$-term is dominant in Eq. \eqref{rho_f} at the time around $N=N_2$, we depict the evolution of $g^2\left(\phi-\phi_0\right)^2\left\langle\chi^2\right\rangle/2$ instead of $\rho_f$ in Fig. \ref{fig2}. Furthermore, we include the evolution of $\Delta \rho$ in the plot as well. It is apparent from the plot that $g^2\left(\phi-\phi_0\right)^2\left\langle\chi^2\right\rangle/2$ exceeds $\Delta\rho$, indicating a violation of the condition \eqref{ECL}. Therefore, the assumption of neglecting the backreaction from the $\chi$ field is an oversimplification. In next section, we will conduct a thorough analysis of $\chi$-particle production, including the effects of backreaction on the evolution of the background and perturbations.

\section{\label{sec3}Backreaction}

In our previous discussion, we have highlighted the potential significance of  the backreaction arising from the produced $\chi$ particles in the evolution of the background and perturbations. Thus, in this section, we carry out numerical computations of the coupled set of evolution equations \eqref{Ein_Eq1_w_Br}-\eqref{omega_k_chi} within Hartree approximation, along with the metric perturbation equation \eqref{fo_pert_Ein_2}.

\begin{figure*}
\centering
\includegraphics[width=1.\textwidth]{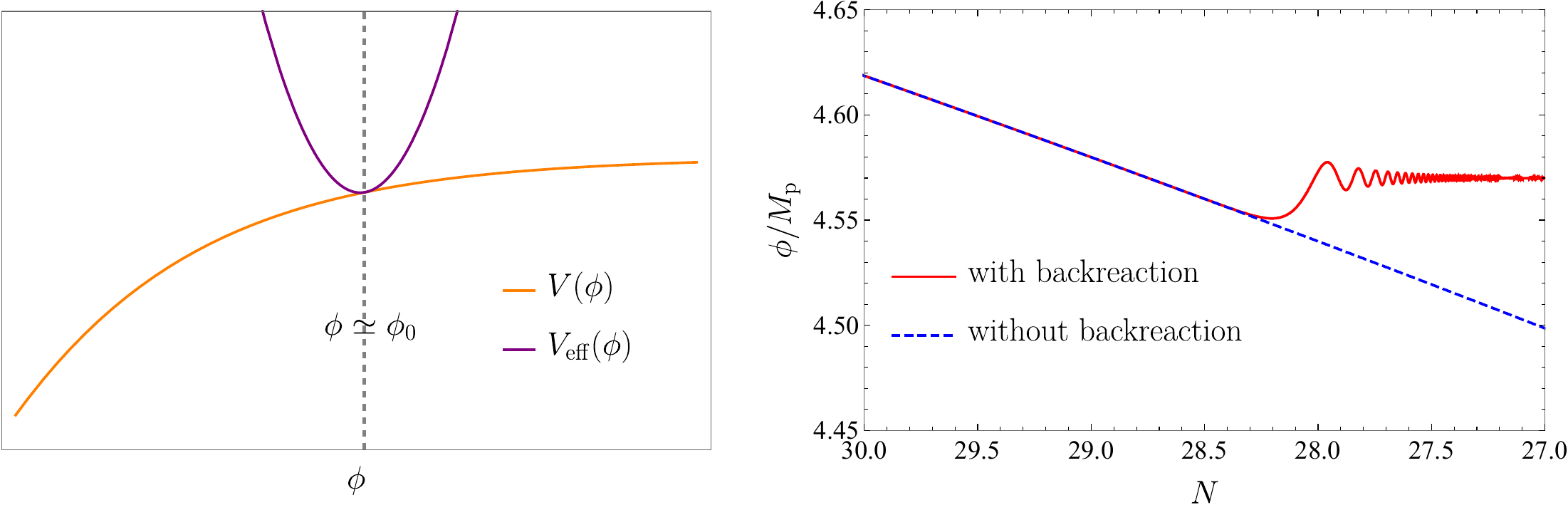}
\caption{{\it Left panel}: The schematic diagram for the inflaton potential $V(\phi)$ and the effective potential of inflaton $V_{\rm eff }(\phi)$ with a sufficient large $\left\langle\chi^2\right\rangle$. {\it Right panel}: The time evolution of $\phi$ with and without the backreaction of the $\chi$ field for the model parameters adopted in Sec. \ref{sec2}.}
\label{fig3}
\end{figure*}

Let us now examine the impact of the substantial amplification of $\chi$ field fluctuations on the background evolution of the inflaton field.
Taking into account the coupling potential between the inflaton and $\chi$ field, the inflaton possesses an effective potential given by
\begin{align}
\label{Eff_Pot}
    V_{\rm eff} = V(\phi) + \frac{1}{2}g^2(\phi-\phi_0)^2\left\langle\chi^2\right\rangle.
\end{align}
Prior to the inflaton approaching the vicinity of $\phi=\phi_0$, that is, prior to the onset of the tachyonic instability, we have $V_{\rm eff} = V(\phi) $ and the inflaton undergoes standard slow-roll evolution. As the inflaton reaches around $\phi_0$, $\left\langle\chi^2\right\rangle$ experience an exponential growth due to the tachyonic instability of specific modes of the $\chi$ field. If $\left\langle\chi^2\right\rangle$ rapidly increases to a sufficiently large value, it causes a transition in the derivative of the effective potential $dV_{\rm eff}/d\phi$ from a positive value to a negative value. Consequently, the global minimum of the effective potential $V_{\rm eff}$ shifts to a field value very close to $\phi_0$, as illustrated in the left panel of Fig. \ref{fig3}. Due to this shift, the inflaton starts oscillating near the minimum of its effective potential, leading to the premature termination of inflation. As a result, inflation is unable to provide a sufficiently large {\it e}-folding number. This situation arises in the model parameter space that leads to a significant tachyonic instability, such as the model parameters selected in the previous section. The right panel of Fig. \ref{fig3} provides a clear illustration of the inflaton's dynamical evolution, considering the backreaction effect for the previously chosen model parameters. It is easy to observe that the inflaton's behavior is consistent with the earlier description, where it enters a phase of oscillations around $\phi_0$. This highlights the importance of taking into account the backreaction of the $\chi$ field when studying the evolution of inflaton.

In this model, the parameter $\xi$ is crucial in determining the strength of the tachyonic instability. A larger value of $\xi$ leads to a more pronounced tachyonic instability. This can result in a more efficient $\chi$-particle production, which can have important implications for the evolution of inflaton. To ensure that the inflation ends in its original way, i.e., through the slow-roll mechanism not dominated by the backreaction effect, an upper bound on the parameter $\xi$ should be imposed. This upper bound ensures that the tachyonic instability is not too strong, preventing the $\chi$ field from dominating the dynamics of the inflaton field. 
The value of the upper bound on $\xi$ depends on the specific inflaton potential and the choice of the parameters $g$ and $\phi_0$. In general, the upper bound can be determined by requiring that the slow-roll conditions are satisfied throughout the inflationary period, and that the backreaction of the $\chi$ field does not lead to a premature end of inflation. For the same value of $g$ and $\phi_0$ as before, the value of the upper bound on $\xi$ has been found to be $\sim 4.16$. 

\begin{figure*}
\centering
\includegraphics[width=1.\textwidth]{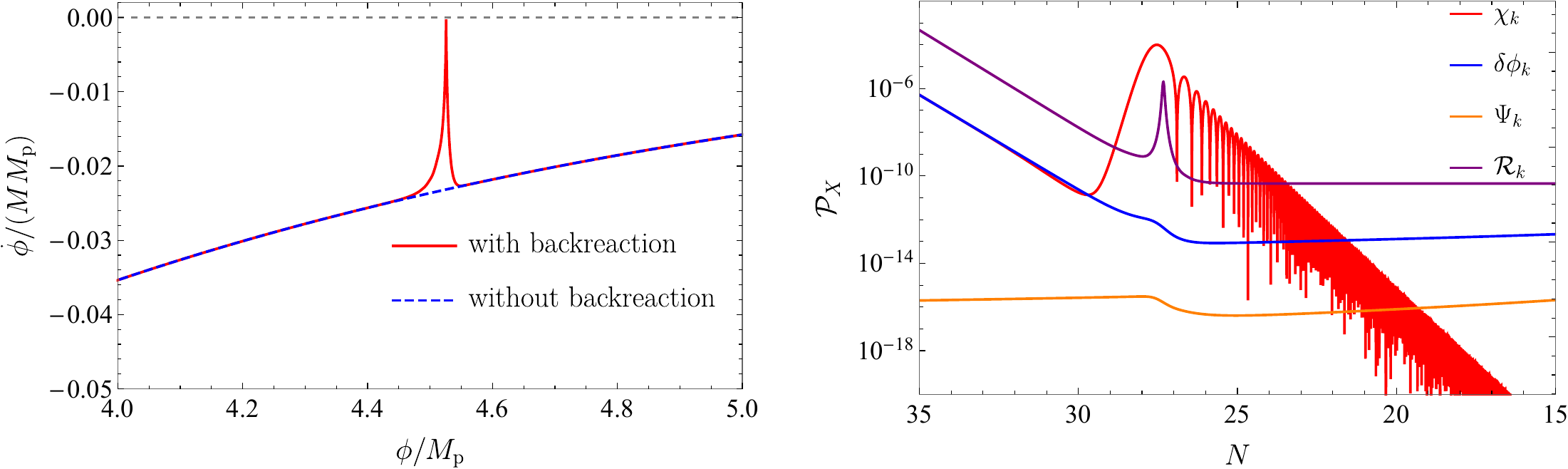}
\caption{{\it Left panel}: The phase portrait in the $\phi-\dot\phi$ plane for the case of $\xi=4.16$ with same $g$ and $\phi_0$ as before. {\it Right panel}: The evolution of $\mathcal{P}_X=k^3/(2\pi^2)|X_k|^2$, for $X_k=\chi_k$, $\delta\phi_k$, $\Psi_k$ and $\mathcal{R}_k$, and using $M_{\rm p}=1$ and the same model parameters as the left panel. This particular $k$ denotes the peak scale of the amplified power spectrum of the $\chi$ field.}
\label{fig4}
\end{figure*}

\begin{figure}
\centering
\includegraphics[width=0.38\textheight]{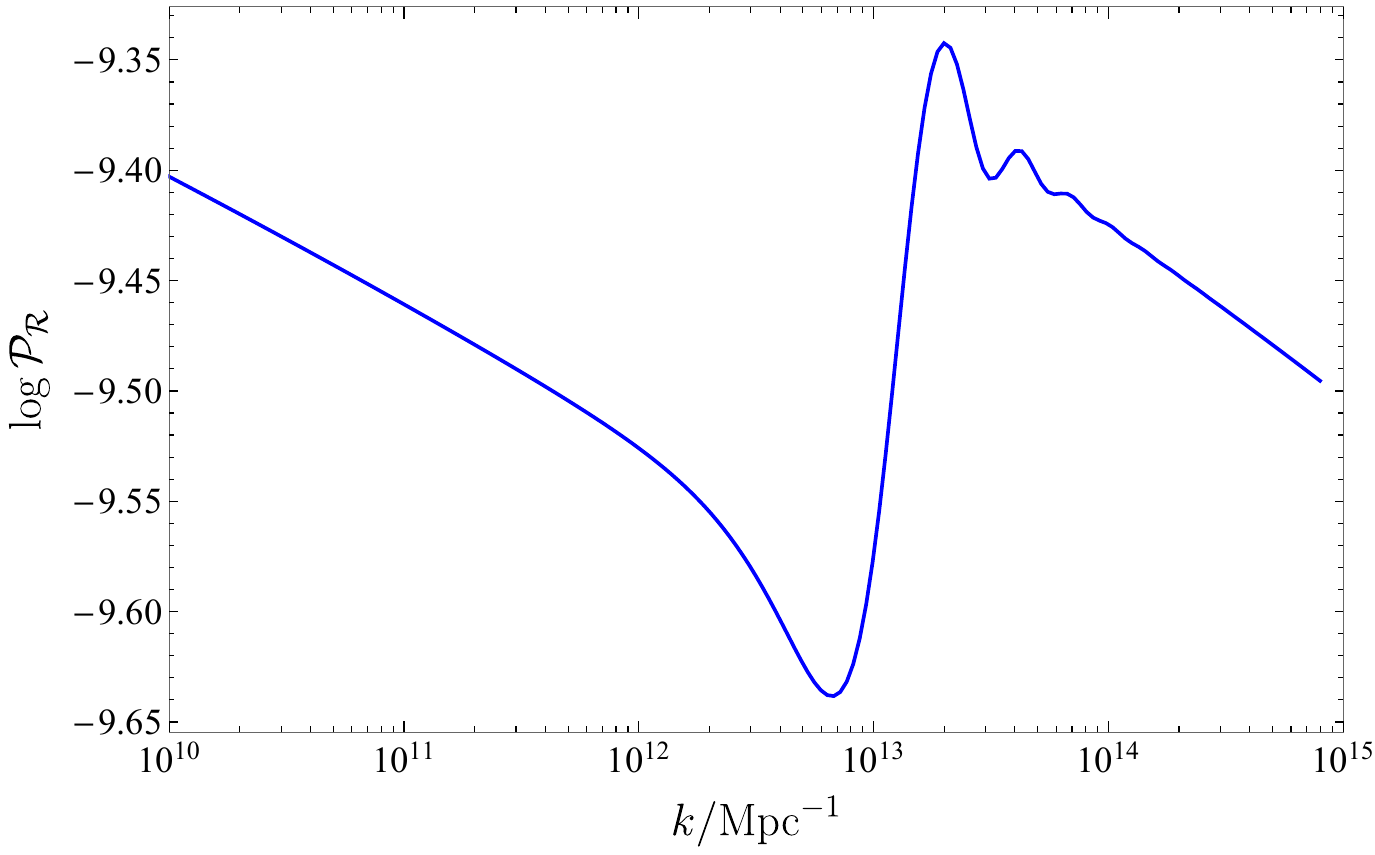}
\caption{The resulting power spectrum of curvature perturbations in the case of considering the backreaction of the $\chi$ field, which is obtained by taking $\phi_0=4.57M_{\mathrm{p}}$, $g = 100 M/M_{\mathrm{p}}$, and $\xi=4.16$.  }
\label{fig5}
\end{figure}

\begin{figure}
\centering
\includegraphics[width=0.38\textheight]{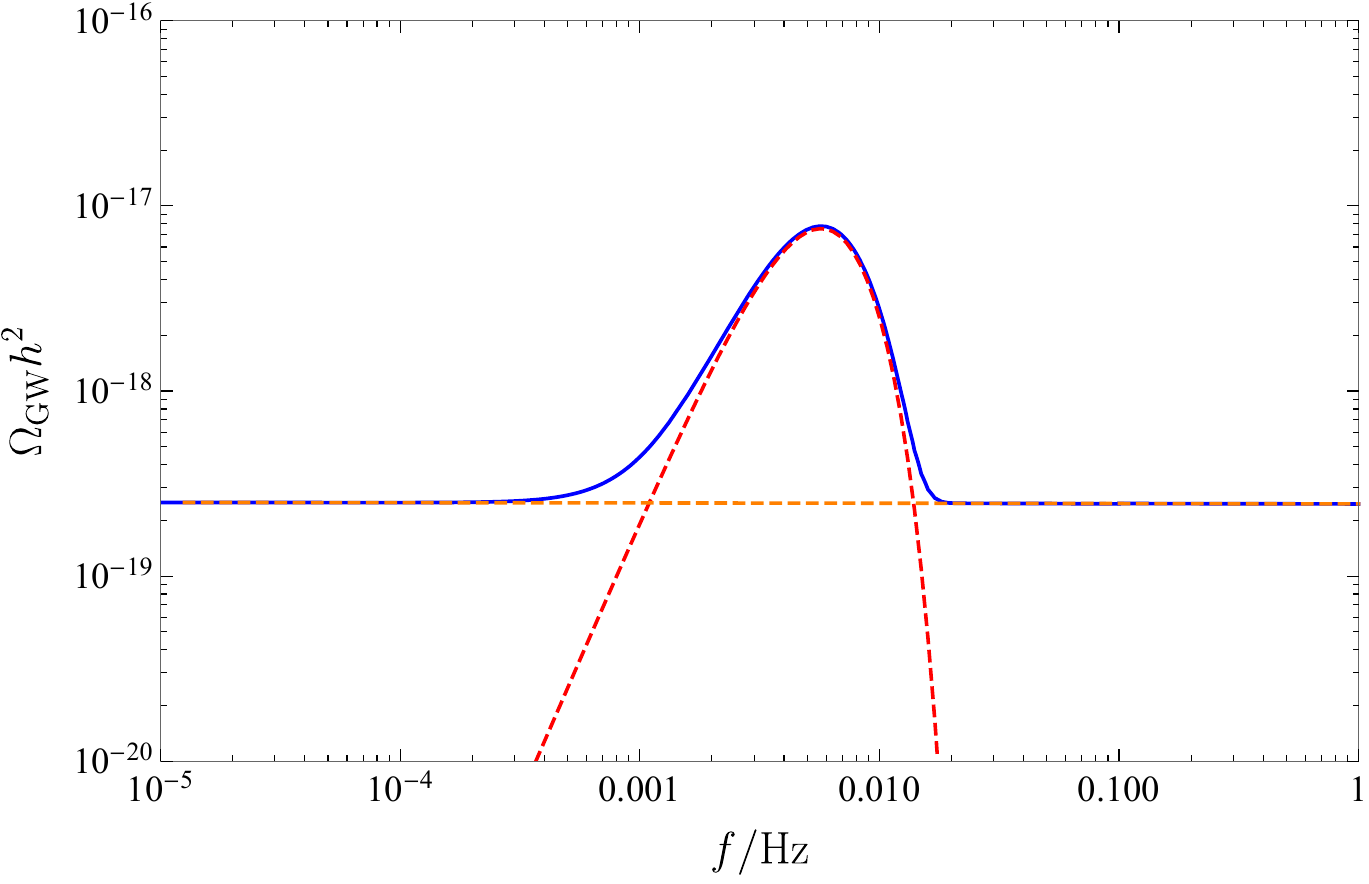}
\caption{ The predicted current energy spectrum of GWs in the case of considering the backreaction of the $\chi$ field, which is obtained by taking $\phi_0=4.57M_{\mathrm{p}}$, $g = 100 M/M_{\mathrm{p}}$, and $\xi=4.16$. The red dashed line represents GWs induced by the amplified $\chi$ field perturbations, and the orange dashed line denotes primordial GWs predicted by the Starobinsky potential. The blue solid line is the total energy spectrum of the resulting GW signal. }
\label{fig6}
\end{figure}

In the left panel of Fig. \ref{fig4}, we show the evolutionary curves in the $\phi-\dot\phi$ plane for the case of $\xi=4.16$ with same $g$ and $\phi_0$ as before. This figure illustrates the dynamics of the inflaton during the inflationary phase when the tachyonic instability of the $\chi$ modes is constrained by the upper bound on $\xi$. As the inflaton field rolls down its potential, it loses all most of its kinetic energy soon after the onset of $\chi$-particle production. However, the tachyonic instability of the $\chi$ modes also come to an end when $\dot\phi \simeq 0$.  At this point, the potential energy of the inflaton, $V(\phi)$, still dominates the energy density of the Universe, ensuring that the inflationary phase continues. As the Universe expands, the expectation value of $\chi^2$, denoted by $\left\langle\chi^2\right\rangle$, quickly decays. This decay leads to a decrease in the coupling potential, which in turn allows the inflaton to return to its slow-roll trajectory. This behavior demonstrates that, even in the presence of tachyonic instability, the inflationary dynamics can be maintained as long as the upper bound on $\xi$ is satisfied. In the right panel of Fig. \ref{fig4}, we present the evolution of the power spectrum at a specific scale for each of the field perturbations $\chi_k$, $\delta\phi_k$, and $\Psi_k$, along with the curvature perturbation $\mathcal{R}_k = \Psi_k + H\delta\phi_k/\dot\phi$. This figure illustrates that the $\chi_k$ mode undergoes an exponential growth and then rapidly decays after reaching its maximum value. As anticipated from perturbation equations \eqref{fo_pert_KG_phi} and \eqref{psi_k}, $\delta\phi_k$ and $\Psi_k$ do not grow due to the absence of a $\chi$ background. Interestingly, the evolution for the $\mathcal{R}_k$ mode exhibits sharp changes before it becomes frozen outside the horizon. This occurs because the friction term in equation of motion for $\mathcal{R}_k$ contains the slow-roll parameter $\eta = \ddot\phi/(H\dot\phi)$, and the drastic changes of $\dot\phi$, as observed in the left panel of Fig. \ref{fig4}, result in significant alterations of $\eta$, which in turn affect the evolution for the $\mathcal{R}_k$ mode. As a consequence, the power spectrum of curvature perturbations for the modes that exit the horizon around the time when there is a significant change in the inflaton velocity will deviate from the near scale invariance expected in the usual slow-roll inflation. In Fig. \ref{fig5}, one can find that the resulting power spectrum of curvature perturbations have the oscillations at these scales. Finally, we present the current energy spectrum of the GW signal predicted by this model in Fig. \ref{fig6}. While the energy spectrum of induced GWs remains significantly distant from the sensitivity curves of the space-based GW experiments, its peak surpasses the amplitude of primordial GWs predicted by the Starobinsky potential by more than an order of magnitude. Consequently, the total energy spectrum of the GW signal exhibits a distinct bump at specific scales. This characteristic could serve as a unique identifier for particle production during inflation, provided that future GW experiments are capable of detecting such subtle GW signal.

\section{\label{sec4}TESTABILITY}
In this section, we explore the testability of the GW signal predicted by this model through the analytical method employed in Refs.~\cite{Cook:2011hg,Kofman:1997yn,Glavan:2015cut,Kim:2021ida}. Firstly, we focus on attaining the analytical evolution of the $\chi$ field with the consideration of neglecting the effect of backreaction. Instead of tackling the Eq. \eqref{EoM_chi} directly, we introduce a new field denoted as $\hat{\chi}_k=a\chi_k$. We then proceed to solve its evolution through the following equation,
\begin{align}
\label{Conformal_EoM_Chi}
\hat{\chi}^{\prime\prime}_k+\hat{\omega}_k^2 \hat{\chi}_k =0,
\end{align}
with
\begin{align}
\hat{\omega}_k^2=k^2+g^2\left(\phi-\phi_0\right)^2a^2-2(6\xi+1)H^2a^2,
\end{align}
where the prime denotes the derivative with respect to the conformal time $\tau$. It is reasonable to assume that the particle production duration is brief, typically expected to be shorter than a Hubble time, thereby ensuring the efficiency of particle production. Therefore, the behavior of the inflaton in the vicinity of $\phi=\phi_0$ can be approximated as a linear function, specifically expressed as $\phi=\phi_0+\dot{\phi}_0(t-t_0)\approx\phi_0-\dot{\phi}_0(\tau/\tau_0-1)/H$, where it is assumed that the Hubble parameter remains constant during inflation, and any quantity with subscript $0$ means it is evaluated when $\phi=\phi_0$. An approximate upper bound on the duration of particle production, represented as $\Delta t \equiv t_e - t_i \simeq 2(t_e-t_0)$ with $t_i$ and $t_e$ being the starting and ending points of the particle production, can be derived by invoking the condition $\hat{\omega}_k^2 < 0$ with $k\simeq 0$. Then, we obtain the following constraint,
\begin{align}
\gamma \equiv \frac{H \Delta t}{2} =\frac{H^2 \sqrt{2(6 \xi+1)}}{-g \dot{\phi}_0}<\frac{1}{2}.
\end{align}
In this case, we can neglect the expansion of the universe during the particle production phase, effectively treating the scale factor $a$ as nearly equivalent to $a_0$. Consequently, Eq. \eqref{Conformal_EoM_Chi} can be reformulated into the Weber equation. By using the initial Bunch-Davies vacuum state, the exact solution assumes the following form,  
\begin{align}
\label{Exact_Sol}
\hat{\chi}_k=2^{-\frac{3}{4}}\left[\frac{1}{\sqrt{\sigma_k}} W\left(-\frac{\kappa^2}{2} ;-\sqrt{2} x\right)+i \sqrt{\sigma_k} W\left(-\frac{\kappa^2}{2} ; \sqrt{2} x\right)\right],
\end{align}
with 
\begin{align}
x=\left(\frac{\tau}{\tau_0}-1\right) \frac{\sqrt{-g \dot{\phi}_0}}{H},  
\end{align}
\begin{align}
\kappa^2=-\frac{k^2 \tau_0^2 H^2}{g \dot{\phi}_0}+\frac{2(1+6 \xi) H^2}{g \dot{\phi}_0},
\end{align}
\begin{align}
\sigma_k=\sqrt{1+e^{-\pi \kappa^2}}-e^{-\frac{\pi \kappa^2}{2}}.
\end{align}
The system transitions back to an adiabatic regime when $t>t_0+\Delta t/2$, and during this period, the solution can be simplified as
\begin{align}
\label{Adiabatic_Exact_Sol}
\hat{\chi}_k=\frac{\alpha_k}{\sqrt{2 \hat{\omega}_k}} \exp \left[-i \int \hat{\omega}_k d \tau\right]+\frac{\beta_k}{\sqrt{2 \hat{\omega}_k}} \exp \left[i \int \hat{\omega}_k d \tau\right],   
\end{align}
The $\alpha_k$ and $\beta_k$ are the Bogoliubov coefficient and obey the relation $|\alpha|^2-|\beta|^2=1$.  Their analytical expression can be determined through the asymptotic expansion of the exact solution \eqref{Exact_Sol} as $\tau\to\infty$, yielding
\begin{align}
\alpha_k=\frac{i}{2}\left(\sigma_k+\frac{1}{\sigma_k}\right), \quad \beta_k=\frac{i}{2}\left(\sigma_k-\frac{1}{\sigma_k}\right).    
\end{align}
The computation of particle number of the field $\hat{\chi}_k$ can be achieved by employing the Bogoliubov coefficients, and it is expressed as
\begin{align}
\left|\beta_k\right|^2=e^{-\frac{\pi\left(k \tau_0\right)^2 H^2}{-g \dot{\phi}_0}} e^{\frac{2 \pi(1+6 \xi) H^2}{-g \dot{\phi}_0}}.    
\end{align}
Obviously, as the parameter $\xi$ increases, the particle number likewise rises. In the limit as $\xi\to\infty$, an unbounded proliferation of particles ensues, leading to the infinite energy density of GWs. Therefore, it is important to impose constraint on the parameter $\xi$ by taking into account the effect of backreaction. In the section \ref{sec3}, we have demonstrated that the maximum energy of perturbations is the kinetic energy of the $\phi$ when the effect of backreaction of the particles reaches its zenith. To analytically estimate the maximum magnitude of the backreaction, we direct our attention to the effective potential \eqref{Eff_Pot} and examine its derivative $dV_{\rm eff}/d\phi$, ensuring its non-negativity, i.e.,
\begin{align}
g^2\left(\phi-\phi_0\right)\left\langle\chi^2\right\rangle + \frac{d V}{d \phi} \geq 0,
\end{align}
to prevent the formation of potential barrier, as exhibited in the left panel Fig. \ref{fig3}. Here, the vacuum expectation value $\left\langle\chi^2\right\rangle\ $ is calculated by Eqs. \eqref{Adiabatic_Exact_Sol} and \eqref{eve}. In the detailed calculation, we have dropped the rapidly oscillating term in the integral and adopted the approximation that $\hat{\omega}_k\approx g\left(\phi_0-\phi\right)a$ when $t>t_0+\Delta t/2$. As a result, the backreaction term has the following form,
\begin{widetext}
\begin{align}
g^2\left(\phi-\phi_0\right)\left\langle\chi^2\right\rangle\approx-\frac{g}{a^3} \int \frac{d k}{2 \pi^2} k^2\left|\beta_k\right|^2=-e^{\frac{2 \pi(1+6 \xi) H^2}{-g \dot{\phi}_0}} \frac{g}{16 \pi^3 a^3} \frac{(-g \dot{\phi}_0)^{\frac{3}{2}}}{H^3 \tau_0^3}.
\end{align}
\end{widetext}
During the period when $t>t_0+\Delta t/2$ but still in the vicinity of $t_0$, the approximation $a\approx a_0$ remains valid. Consequently, we can deduce an upper limit for the parameter $\xi$ through the following equation,
\begin{align}\label{Constrain_xi}
e^{\frac{2 \pi(1+6 \xi) H^2}{-g \dot{\phi}_0}}=\frac{d V}{d \phi} \frac{16 \pi^3}{g(-g \dot{\phi}_0)^{\frac{3}{2}}}.    
\end{align}

Finally, we employ the methodology and assumptions outlined in Ref.~\cite{Cook:2011hg} to evaluate the power spectrum of gravitational waves arising from particle production. This power spectrum is expressed as follows:
\begin{widetext}
\begin{align}
\mathcal{P}_h \left(k\right)=\frac{2 H^2}{\pi^2 M_{\mathrm{p}}^2} \frac{H^2}{\pi^5 M_{\mathrm{p}}^2}\left(\frac{-g \dot{\phi}_0}{H^2}\right)^{\frac{3}{2}} e^{\frac{2 \pi(1+6 \xi) H^2}{-g \dot{\phi}_0}}\left(1+\frac{1}{8 \sqrt{2}} e^{\frac{2 \pi(1+6 \xi) H^2}{-g \dot{\phi}_0}}\right) \frac{F\left(k \tau_0\right)_\gamma}{\left(k \tau_0\right)^3},    
\end{align}
\begin{align}
 F\left(k \tau_0\right)_\gamma \approx\left[k \tau_0 \cos \left(k \tau_0\right)-\sin \left(k \tau_0\right)\right]^2(\ln \gamma)^2.
\end{align}
\end{widetext}
The term $F\left(k \tau_0\right)_\gamma/\left(k \tau_0\right)^3$ reaches its maximum value when $k\tau_0\approx2.5$, signifying the emergence of the peak in the power spectrum of GWs. Introducing the slow-roll paramter $\epsilon=\dot\phi^2/(2H^2M_{\mathrm{p}}^2)\simeq (M_{\mathrm{p}}^2/2)[(dV/d\phi)/(3H^2M_{\mathrm{p}}^2)]^2$, the peak value of the power spectrum of this GW signal can be written as
\begin{align}
\mathcal{P}_h \simeq \frac{329.3\epsilon_0^{\frac{1}{4}}(\ln \gamma)^2}{g^{\frac{7}{2}}}\left(\frac{H}{M_{\mathrm{p}}}\right)^{\frac{3}{2}}\mathcal{P}^{\rm (p)}_h,
\end{align}
where $\mathcal{P}^{\rm (p)}_h=(2 H^2)/(\pi^2 M_{\mathrm{p}}^2)$ denotes the power spectrum of the primordial GWs. The maximum value of $H/M_{\mathrm{p}}$ can be estimated to be $2\times10^{-5}$ through the upper bound on tensor-to-scalar ratio and the amplitude of the curvature power spectrum at CMB scale. Furthermore, in virtue of the Eq. \eqref{Constrain_xi} and the Eq. \eqref{ES_GW}, we can estimate the energy spectrum of the resulting GWs as follows:
\begin{equation}\label{Analy_ES_GW}
\Omega_{\mathrm{GW},0}(k) < 3.785\times 10^{-22} \frac{\epsilon_0^{\frac{1}{4}}(\ln \gamma)^2}{g^{\frac{7}{2}}},
\end{equation}
with
\begin{equation}\label{Para_time}
 \gamma= \frac{3\times 10^{-3}}{\epsilon_0^{\frac{1}{4}} g^{\frac{1}{2}}} \ln^{\frac{1}{2}}\left( \frac{5.58}{g^{\frac{5}{2}} \epsilon_0^{\frac{1}{4}}}\right) < \frac{1}{2}.
\end{equation}
Given that the particle production occurs during the slow-roll inflation, $\epsilon_0$ must at least meet the condition $\epsilon_0 \lesssim 0.1$. Combining this condition with the constraint provided in Eq.~\eqref{Para_time}, the maximum value of $\epsilon_0^{1/4}(\ln \gamma)^2/g^{7/2}$ can be achieved at $g=0.009181$ and $\epsilon_0=0.05$.
Consequently, this establishes an upper limit on the GW energy spectrum, specifically, $\Omega_{\mathrm{GW},0}(k)<4.6\times10^{-15}$, which is two orders of magnitude greater than the primordial one. This upper bound remains below the detection limits of LISA and Taiji but slightly exceeds the sensitivity curves of BBO \cite{Phinney:2003} and SKA.
However, it's important to note that the analytically derived upper limit on the GW energy spectrum may be an overestimate. In practice, our numerical results reveal that in the context of typical inflationary potentials, such as the Starobinsky potential, power law potential, and axion potential, the maximum peak value of the resulting GW energy spectrum is on the order of  $10^{-16}$, which is unobservable by any prospective GW experiment.
Nevertheless, there remains the potential for detecting this signal through GW experiments like BBO and SKA, provided that we meticulously design the special inflationary potentials within our model.

\section{Conclusion and Discussion}
\label{sec5}
In this article, we explore the phenomenology of particle production for a spectator scalar field $\chi$ during inflation. The $\chi$ field is assumed to be coupled to both the inflaton and the Ricci scalar. The interaction between the $\chi$ field and gravity can cause the effective mass square of the $\chi$ field to become negative, which in turn triggers the tachyonic instability of specific $\chi$ modes. As a result, the amplified $\chi$ field will act as a GW source, generating a GW signal. If we disregard the backreaction of the $\chi$ field and select suitable model parameters, particularly the $\xi$ value that is positively correlated with the strength of the tachyonic instability, specific modes of the $\chi$ field will be significantly amplified, making the resulting GW signal detectable by LISA and Taiji. However, in reality, the backreaction of the $\chi$ field will cause the premature termination of inflation in this case.

To guarantee that the inflation ends via the slow-roll mechanism, it is necessary to impose an upper bound on the parameter $\xi$. In the case of adopting this upper bound and taking into account the backreaction of the $\chi$ field, we observe that the inflaton velocity almost approaches zero shortly after the emergence of the tachyonic instability, however it quickly reverts to the slow-roll regime. This evolution of the inflaton results in a special oscillating structure in the power spectrum of curvature perturbations at certain scales. Furthermore, considering the upper limit of the parameter $\xi$, the energy spectrum of GWs induced by the $\chi$ field is analytically estimated to be $\Omega_{\mathrm{GW},0}(k)<4.6\times10^{-15}$. Despite this estimate may tend to be on the higher side, the GW signal remains unobservable by LISA and Taiji. In the case of specific inflationary potentials, such as the Starobinsky potential, power law potential, and axion potential, our numerical results indicate that the amplitude of GW energy spectrum is typically on the order of $10^{-16}$, which is below the detection limit of any prospective GW experiment. However, if we meticulously design the unique inflationary potentials within our model, there still exists the prospect of detecting the predicted GW signal through experiments like BBO and SKA. We can delve into this possibility in future discussions.

Note that we focus on the phenomenology at small scales in this paper. Interestingly, if we shift our attention to the CMB scales by take the value of the parameter $\phi_0$ close to the initial field value of the inflaton, the phenomena predicted in this model can lead to some intriguing results. On one hand, an appropriate scalar power spectrum with superimposed oscillations could potentially explain the large scale CMB anomalies as discussed in recent works \cite{Braglia:2021sun,Tiwari:2022zzz}. On the other hand, the presence of an induced component in the total GW signal would violate the standard consistency relation between the tensor-to-scalar ratio and tensor spectra index, as predicted by the usual single-field slow-roll inflation. These topics are fascinating and warrant further investigation in future studies.

\begin{acknowledgments}
We thank Zhi-Zhang Peng for useful discusstions. This work is supported in part by the National Key Research and Development Program of China Grant No. 2020YFC2201501, in part by the National Natural Science Foundation of China under Grants No. 12075297, No. 12235019, and No. 12305057.
\end{acknowledgments}

\appendix

\section{\label{A}The basic equations with backreaction of perturbations}
In this appendix, we derive the basic equations with backreation of perturbations by following the reference~\cite{Zibin:2000uw}. It is common to separate the inflaton field into a homogeneous background $\phi$, and a perturbation $\delta\phi$. In the case of the $\chi$ field, it is not necessary to make a similar separation because it is already considered as a quantum field.
Throughout this paper, we work with the spatially flat FRW metric in the conformal Newtonian gauge, and then the perturbed metric, incorporating both the first-order scalar metric perturbation $\Psi$ and the second-order tensor perturbation $h_{ij}$, can be written as
\begin{align}
\label{Metric}
d s^2=-\left(1+2\Psi\right)dt^2+a^2\left[\left(1-2\Psi\right) \delta_{i j}+\frac{h_{i j}}{2}\right]d x^i d x^j.
\end{align}
A common method for approximately estimating the impact of field fluctuations on the background and perturbation evolution is to incorporate Hartree terms into the equation of motion \cite{Kofman:1997yn}. In this approximation, the background equations are as follows:
\begin{widetext}
\begin{align} \label{Ein_Eq1_w_Br} 
3M_{\mathrm{p}}^2H^2 =& \frac{1}{1+\xi M_{\mathrm{p}}^{-2} \left\langle\chi^2\right\rangle}\left[\frac{1}{2}\dot{\phi}^2+\frac{1}{2}\left\langle\delta \dot{\phi}^2\right\rangle+\frac{1}{2}\left\langle\dot{\chi}^2\right\rangle+\frac{1}{2 a^2}\left\langle (\nabla\delta\phi)^2\right\rangle+\frac{1}{2 a^2}\left\langle(\nabla\chi)^2\right\rangle \right. \nonumber \\
&\left. -6\xi H\left\langle \chi\dot\chi \right\rangle+ \frac{\xi}{a^2}\left\langle\nabla^2(\chi^2) \right\rangle+V(\phi)+\frac{1}{2}\frac{\partial^2 V }{\partial \phi^2}\left\langle\delta\phi^2 \right\rangle+\frac{1}{2}g^2\left(\phi-\phi_0\right)^2\left\langle\chi^2 \right\rangle \right],
\end{align}
\end{widetext}
\begin{widetext}
\begin{align} \label{Ein_Eq2_w_Br}
 -M_{\mathrm{p}}^2\left(2\dot{H}+3H^2\right)= & \frac{1}{1+\xi M_{\mathrm{p}}^{-2} \left\langle\chi^2\right\rangle} \left[ \frac{1}{2} \dot{\phi}^2 +\frac{1}{2}\left\langle\delta \dot{\phi}^2\right\rangle +\left( \frac{1}{2} +2\xi\right)\left\langle\dot{\chi}^2\right\rangle - \frac{1}{6a^2}\left\langle(\nabla\delta\phi)^2\right\rangle  \right. \nonumber \\
& \left. - \frac{1}{6a^2} \left\langle(\nabla\chi)^2\right\rangle - \frac{2\xi}{3 a^2} \left\langle\nabla^2(\chi^2)\right\rangle + 4\xi H\left\langle \chi\dot\chi\right\rangle + 2\xi\left\langle \chi\ddot\chi\right\rangle - V(\phi) \right. \nonumber \\
&\left.- \frac{1}{2} \frac{\partial^2 V}{\partial \phi^2}\left\langle\delta \phi^2\right\rangle - \frac{1}{2} g^2\left(\phi-\phi_0\right)^2\left\langle\chi^2\right\rangle \right],
\end{align}
\begin{align} \label{KG_Eq_w_Br}
\ddot{\phi}+3 H \dot{\phi}+\frac{\partial V}{\partial \phi}+\frac{1}{2}\frac{\partial^3 V}{\partial \phi^3}\left\langle\delta \phi^2\right\rangle+g^2\left(\phi-\phi_0\right)\left\langle\chi^2\right\rangle=0, 
\end{align}
\end{widetext}
where the notation $\left\langle...\right\rangle$ represents the expectation value, calculated by e.g.
\begin{align}\label{eve}
    \left\langle\chi^2\right\rangle = \frac{1}{(2\pi)^3} \int d^3k |\chi_k|^2.
\end{align}
Then, the momentum-space linearized field perturbation equations are given by
\begin{align}
\label{fo_pert_KG_phi}
\delta\ddot{\phi}_k+3H \delta\dot{\phi}_k+\Omega_k^2 \delta\phi_k =4\dot{\Psi}_k\dot{\phi}-2\frac{\partial V}{\partial \phi}\Psi_k,
\end{align}
with
\begin{align}
\Omega_k^2=\frac{k^2}{a^2}+\frac{\partial^2 V}{\partial\phi^2}+g^2\left\langle\chi^2\right\rangle+\frac{1}{2}\frac{\partial^4 V}{\partial \phi^4}\left\langle\delta\phi^2\right\rangle,
\end{align}
and
\begin{align}
\label{EoM_chi_k_w_Br}
\ddot{\chi}_k+3H \dot{\chi}_k+\omega_k^2  \chi_k =0,
\end{align}
with
\begin{align}\label{omega_k_chi}
\omega_k^2=\frac{k^2}{a^2}+g^2\left(\phi-\phi_0\right)^2+g^2\left\langle\delta\phi^2\right\rangle-6\xi\left(\dot{H}^2+2H^2\right),
\end{align}
where $\Psi_k$ obeys following perturbed Einstein equations,
\begin{widetext}
\begin{align}
\label{fo_pert_Ein_1}
3H\dot{\Psi}_k + \left(  \frac{k^2}{a^2} + 3H^2 \right) \Psi_k = - \frac{1}{2 M_\mathrm{p}^2} \left( \dot\phi\delta\phi_k - \Psi_k\dot\phi^2 + \frac{dV}{d\phi}\delta\phi_k \right),
\end{align}
\end{widetext}
\begin{align}
\label{fo_pert_Ein_2}
\dot{\Psi}_k + H \Psi_k =\frac{\dot{\phi} }{2 M_\mathrm{p}^2} \delta \phi_k.
\end{align}
Equations \eqref{fo_pert_Ein_1} and \eqref{fo_pert_Ein_2} can be combined to give
\begin{align}
\label{psi_k}
\Psi_k=\frac{\dot{\phi} \delta \dot{\phi}_k+3H\dot{\phi} \delta \phi_k+(\partial V/\partial \phi) \delta \phi_k}{\dot{\phi}^2- 2 M_{\mathrm{p}}^2\left(k/a\right)^2},
\end{align}

Next, we derive the equation of motion for the second-order tensor perturbation $h_{ij}$, given by
\begin{align}
\label{EoM_IGW}
h_{i j}^{\prime\prime}(\tau,\boldsymbol{x})+2 \mathcal{H}h_{i j}^\prime(\tau,\boldsymbol{x}) -\nabla^2 h_{i j}(\tau,\boldsymbol{x})=\frac{4}{M_{\mathrm{p}}^2} \pi_{i j}^{TT}(\tau,\boldsymbol{x}),
\end{align}
where a prime denotes the derivative with respect to the conformal time $\tau\equiv \int^t dt/a$, and $\mathcal{H}\equiv a^{'}/a$ denotes the conformal Hubble parameter. The source term on the right side of the equation can be written as
\begin{align}
\label{GWSourceTerm}
\pi_{i j}^{T T}=\hat\pi_{i j}^{l m} \left[(1+2\xi)\partial_l \chi \partial_m \chi+2\xi\chi\partial_m \partial_l \chi\right],
\end{align}
where $\hat\pi_{i j}^{l m}$ is the transverse-traceless projection operator. 
It should be noted that the contribution of the $\phi$ field perturbations to the GW source term is disregarded in light of their lack of enhancement in this model.
By virtue of the polarization tensor $e^\lambda_{i j}$ with $\lambda=+,\times$, we can expand the tensor perturbation and the source term in Fourier space, respectively, as 
\begin{align}
h_{i j}= \sum\limits_{\lambda=+,\times} \int \frac{d \boldsymbol{k}^3}{(2\pi)^{3/2}}e^{i\boldsymbol{k}\cdot\boldsymbol{x}}e^\lambda_{i j}(\boldsymbol{k})h_k^{\lambda}(\tau),
\end{align}
\begin{widetext}
\begin{align}
\pi_{i j}^{T T} &= \sum\limits_{\lambda=+,\times} \int \frac{d \boldsymbol{k}^3}{(2 \pi)^{3/2}} e^{i \boldsymbol{k} \cdot \boldsymbol{x}}e^\lambda_{i j} e^{\lambda,l m} \int \frac{d \boldsymbol{p}^3}{(2 \pi)^{3/2}} \chi_{|\boldsymbol{k}-\boldsymbol{p}|} \chi_p\left[(2\xi+1) \left(p_l p_m - k_l p_m \right)-2\xi p_l p_m \right]\nonumber \\
&= \sum\limits_{\lambda=+,\times} \int \frac{d \boldsymbol{k}^3 d \boldsymbol{p}^3}{(2 \pi)^3} e^{i \boldsymbol{k} \cdot \boldsymbol{x}}e^\lambda_{i j} e^{\lambda,l m} p_l p_m \chi_{ | \boldsymbol{k}-\boldsymbol{p} |} \chi_p.
\end{align}
\end{widetext}
By putting above formulas substitute into Eq. \eqref{EoM_IGW}, we have
\begin{align}
h_k^{\lambda \prime \prime}+2\mathcal{H} h_k^{\lambda\prime}+k^2 h^\lambda_k=\frac{4 }{M_p^2} \int \frac{d^3 \boldsymbol{p}}{(2\pi)^{3/2}}e^{\lambda,ij}p_ip_j \chi_{|\boldsymbol{k}-\boldsymbol{p}|}\chi_{p}.
\end{align}
We solve this equation through the Green’s function method, i.e.,
\begin{align}
h^\lambda_k(\tau) = \frac{4 }{M_{\mathrm{p}}^2} \int^\tau d\tau^\prime G_k(\tau,\tau^\prime) \int \frac{d^3 \boldsymbol{p}}{(2\pi)^{3/2}}e^{\lambda,ij}p_ip_j \chi_{|\boldsymbol{k}-\boldsymbol{p}|}\chi_{p},
\end{align}
where $G_k(\tau,\tau^\prime)$ is the Green’s function. The Starobinsky potential is flat enough to use the de-Sitter approximation to get the Green's function, so that $a=-1/(H\tau)$ and the Green’s function reads \cite{Biagetti:2013kwa}
\begin{widetext}
\begin{align}
G_k\left(\tau, \tau^\prime\right)= \frac{1}{k^3\tau^{\prime 2}}\left[\left(1+k^2 \tau \tau^\prime\right) \sin k(\tau-\tau^\prime)-k(\tau-\tau^\prime) \cos k(\tau-\tau^\prime)\right] \Theta(\tau-\tau^\prime).
\end{align}
\end{widetext}
After some algebraic operation. The power spectrum, $\mathcal{P}_h = \sum_{\lambda=+,\times}k^3/(2\pi^2)|h_k^\lambda|^2$, can be calculated as 
\begin{align}
\label{PS_GW}
\mathcal{P}_h(k)=\frac{2 k^3}{\pi^4 M_{\mathrm{p}}^4} \int_0^{\pi} d \theta \sin ^5 \theta \int d p p^6\left|\int^{\tau} d \tau^\prime G_k\left(\tau, \tau^\prime\right)\chi_p\chi_{|\boldsymbol{k}-\boldsymbol{p}|}\right|^2.
\end{align}
The current energy spectrum of GWs is related to the power spectrum of tensor perturbations through~\cite{Inomata:2021zel}
\begin{align}\label{ES_GW}
\Omega_{\mathrm{GW}, 0}(k) \simeq 1.7 \times 10^{-7} \mathcal{P}_h\left(k\right).
\end{align}

\bibliography{main_}

\end{document}